\documentclass{iaus}
\usepackage{graphicx}

\title[S 262.~~ Probing the stellar population of seyfert galaxies] 
{Probing the stellar population of seyfert galaxies: a near infrared perspective}

\author[Riffel et al.]   
{Rog\'erio Riffel$^1$; Miriani G. Pastoriza$^1$; Alberto Rodr\'{\i}guez-Ardila$^2$
 \and Charles Bonatto$^1$}

\affiliation{$^{1}$Departamento de Astronomia, Universidade Federal do Rio Grande 
do Sul. Av. Bento Gon\c calves 9500, Porto Alegre, RS, Brazil.\\ email: {\tt riffel, miriani.pastoriza, charles.bonatto @ufrgs.br} \\[\affilskip]
$^2$ Laborat\'{o}rio Nacional de Astrof\'{i}sica/MCT - Rua dos Estados Unidos 154, Bairro das Nac\~oes, Itajub\'a, MG, Brazil.
\\email: {\tt aardila@lna.br}}

\pubyear{2009}
\volume{262}  
\pagerange{1--4}
\jname{Stellar Populations - Planning for the Next Decade}
\editors{Gustavo Bruzual \& Stephane Charlot, eds.}
\begin{document}

\maketitle

\begin{abstract}

We employ IRTF SpeX NIR (0.8--2.4$\mu$m) spectra to investigate the stellar population
(SP), active galactic nuclei (AGN), featureless continuum ($FC$) and hot dust properties
in 9 Sy\,1 and 15 Sy\,2 galaxies. Both the {\sc starlight} code and the hot dust as an
additional base element were used for the first time in this spectral range.  Our synthesis shows 
significant differences between Sy\,1 and Sy\,2
galaxies: the hot dust component is required to fit the $K$-band spectra of $\sim$90\%
of the Sy\,1 galaxies, and only of $\sim$25\% of the Sy\,2; about 50\% of the Sy\,2 galaxies
require an $FC$ component contribution $\gtrsim$20\%; this fraction increases
to about 60\% in the Sy\,1. In about 50\% of the Sy2, the combined FC and young components
contribute with more than 20\%, while this occurs in 90\% of the Sy1, suggesting recent
star formation in the central region. The central few hundred parsecs of our galaxy sample
contain a substantial fraction of intermediate-age SPs with a mean metallicity near solar. Our SP 
synthesis confirms that the 1.1$\mu$m CN band can be used as a tracer of intermediate-age
stellar populations. 
\keywords{galaxies: active, galaxies: stellar content, stars: AGB and post-AGB.}
\end{abstract}

\firstsection 
\section{Introduction}

To determine if circumnuclear stellar populations (SPs) and nuclear activity are closely 
related phenomena, or if they are only incidental, it is of utmost importance the correct characterisation 
of the former, since a substantial fraction of the energy emitted 
by a galaxy in the optical to near-infrared (NIR) domain is starlight. 
Moreover, the analysis of the stellar content provides information on 
critical processes such as the star formation episodes and the 
evolutionary history of the galaxy. 

One reason to use the NIR to study the SP of AGNs is that it is the most convenient
spectral region accessible to ground-based telescopes to probe highly obscured sources. 
However, tracking the star formation in the NIR is complicated (\cite[Origlia \& Oliva 2000]{origlia00}). Except for 
a few studies such as those based on the Br$\gamma$ emission or the CO(2-0)
first overtone (\cite[e.g. Origlia et al.  1993]{orig93}), the SP
of the inner few hundred parsecs of active galaxies in the NIR remains poorly known 
(\cite[Riffel et al 2008, 2009]{riffel08,riffel09}).
Because stellar absorption features in the NIR are widely believed to
provide a means for recognizing red supergiants (\cite[Oliva et al. 1995]{oliva95}), they
arise as prime indicators for tracing starbursts in galaxies.
Besides the  short-lived red supergiants, the NIR also includes 
the contribution of thermally- pulsating asymptotic giant
branch (TP-AGB) stars, enhanced in young to intermediate age stellar
populations (\cite[$0.2 \leq t \leq 2$ Gyr, Maraston 2005]{maraston05}).
The TP-AGB phase becomes fully developed 
in stars with degenerate carbon oxygen cores (\cite[see Iben \& Renzini 1983, for a review]{ir83}).
 Evidence of this population in the optical is usually missed, as the most
prominent spectral features associated with this population fall in the NIR \cite{maraston05}. 

With the new generations of Evolutionary Population Synthesis (EPS) models, which 
include a proper treatment of the TP-AGB phase (\cite[Maraston 2005]{maraston05}), it is now possible 
to study the NIR SP of galaxies in more detail. According
to these models, the effects of TP-AGB stars in the NIR spectra are unavoidable. 
\cite[Maraston (2005)]{maraston05} models, by including empirical 
spectra of oxygen-rich stars (\cite[Lan\c con \& Wood 2000]{lw00}), are able to foresee the presence of NIR 
absorption features such as the 1.1$\mu$m  CN band (\cite[Riffel et al. 2007]{riffel07}), whose  
detection can be taken as an unambiguous evidence of a young to intermediate age stellar population.

\section{Spectral Synthesis}

Clearly, the most important ingredient in the SP synthesis is the spectral base 
set, $b_{j,\lambda}$. An ideal base of elements should cover the range of spectral properties observed 
in the galaxy sample, providing enough resolution in age and metallicity to properly address the desired
scientific question (\cite[Schmidt et al. 1991,Cid Fernandes et al. 2005]{alex91,cid05a}).

One improvement here over previous approaches that attempted to describe the stellar content
of active galaxies using NIR spectroscopy is the inclusion of EPS models that
take into account the effects of TP-AGB stars. Accordingly, we use as base set the EPS of 
\cite[Maraston 2005]{maraston05}. The SSPs used in this work
cover 12 ages, $t$= 0.01, 0.03, 0.05, 0.1, 0.2, 0.5, 0.7, 1, 2, 5, 9, 13\,Gyr, and 4 metallicities, namely: 
$Z$= 0.02\,$Z_\odot$, 0.5\,$Z_\odot$, 1\,$Z_\odot$ and 2\,$Z_\odot$, summing up 48 SSPs.

When trying to describe the continuum observed in AGNs, the signature
of the central engine cannot be ignored. Usually, this component is represented by a featureless continuum 
(\cite[$FC$, e.g.Cid Fernandes et al. 2004 and references therein]{cid04}) of power-law form that 
follows the expression $F_{\nu}\propto \nu^{-1.5}$. Therefore, this component was also
added to the base of elements. In the spectral region studied here, hot dust plays an important 
role in the continuum emission of active galaxies.  
Previous studies  (\cite[i.e., Riffel et al. 2006, for instance]{riffel06}) report a minimum in the continuum emission 
around 1.2$\mu$m, probably associated with the red end of the optical continuum related to the central 
engine and the onset of the emission due to reprocessed nuclear radiation by dust 
(\cite[Riffel et al 2009 and references therein]{riffel09}).  In order to properly account for this component,
we have included in our spectral base 8 Planck distributions (black-body-$BB$), 
with $T$ ranging from 700 to 1400\,K, in steps of 100\,K.

\section{Results}

In this work we investigate the NIR spectra of 24 Seyfert galaxies (9 Sy\,1 and 15 Sy\,2) observed
with the IRTF SpeX, obtained in the short cross-dispersed mode. The results of the spectral synthesis fitting procedure 
are presented and discussed in details in \cite[Riffel et al. (2009)]{riffel09}. The approach followed 
here is based on the {\sc starlight} code (\cite[Cid Fernandes et al. 2004,2005, and references therein]{cid05}), which considers the whole observed spectrum, continuum
and absorption features.

\begin{figure*}
\begin{minipage}[b]{0.5\linewidth}
\includegraphics[width=\textwidth]{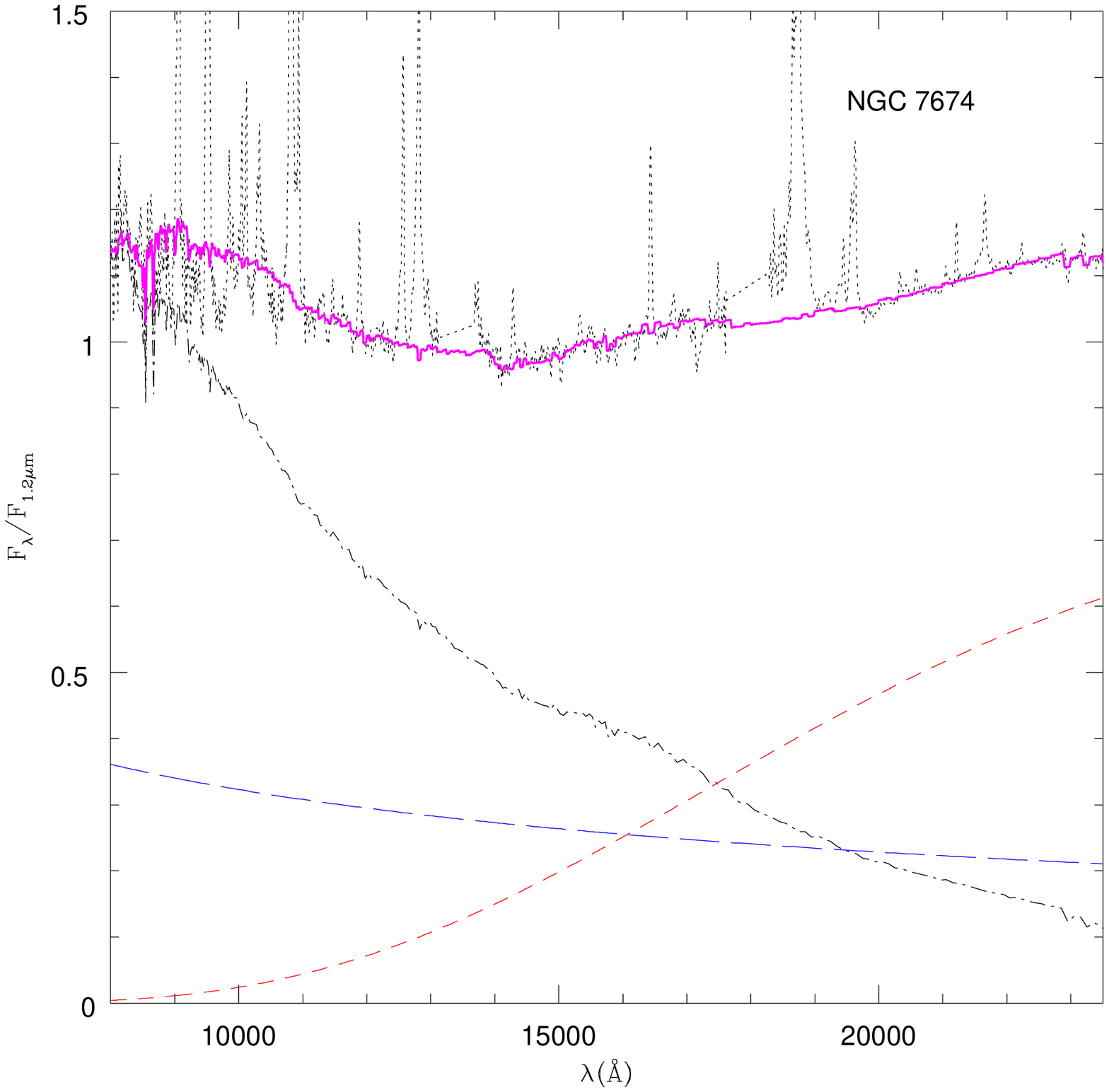}
\end{minipage}\hfill
\begin{minipage}[b]{0.5\linewidth}
\includegraphics[width=\textwidth]{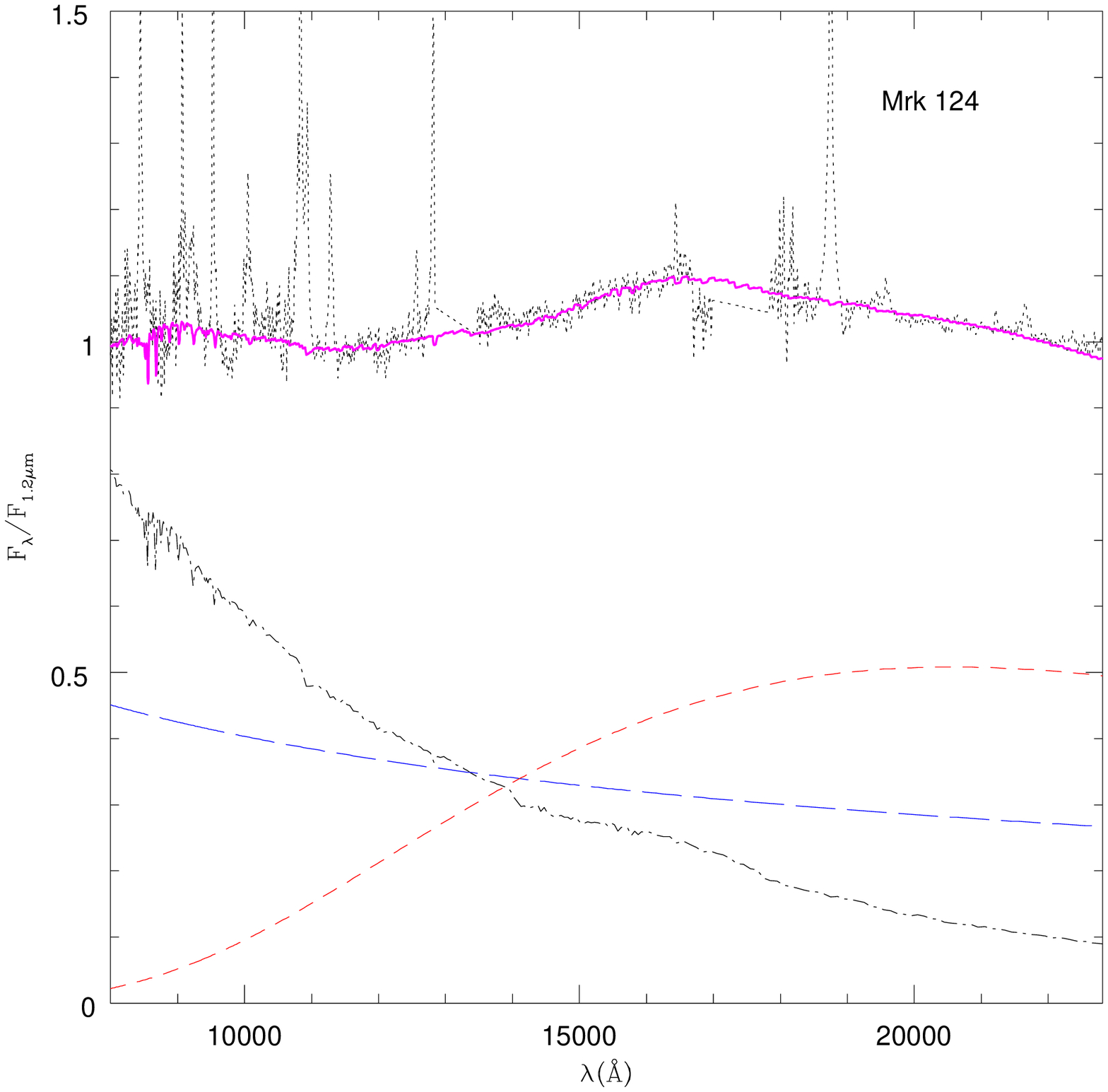}
\end{minipage}\hfill
\caption[]{Three continuum components of NGC\,7674 and Mrk\,124. Dot-short dashed line represents the stellar population 
($x_{\rm y} + x_{\rm I} + x_{\rm o})$. The FC and hot dust component are represented 
by the long and short dashed lines, respectively. The solid line is the sum of the three components and the 
dotted line represents the observed spectrum.}
\label{3comp}
\end{figure*}

In genneral, the spectral synthesis shows that the NIR continuum of active galaxies 
can be explained in terms of at least three components: 
a non-thermal continuum, the dust emission and the stellar population of the circumnuclear region. As can be seen in Fig.~\ref{3comp},  the contribution of 
the latter to the nuclear continuum is higher than 50\% in most objects.  Therefore, its study is a critical step in the analysis of the continuum 
emission of Seyfert galaxies.   Moreover, our results are consistent with the predictions of the unified model for AGNs, 
as the non-thermal continuum and the hot dust emission are present in all Sy\,1 sources and only in a 
small fraction of the Sy~2s.

Note that to take into account noise effects that dump small differences between 
similar spectral components, we present our results using a condensed population vector, which is obtained by binning
the $\vec{x}$ into {\bf young}, $x_Y$ ($t_j\leq \rm 5\times 10^7$yr); {\bf intermediate-age}, 
$x_I$ ($1\times 10^8 \leq t_j\leq \rm 2\times 10^9$yr) and {\bf old}, $x_O$ ($t_j > \rm 2\times 10^9$yr) components, 
using the flux contributions.  We have also binned 
the black-body  contributions into two components. The cool ($BB_c$) is obtained by summing 
up the $BB$ contributions with T $\leq$ 1000\,K, and  the hot one ($BB_h$) with T$\geq$ 1100\,K. For more details on vector 
definition see \cite[Riffel et al. (2009)]{riffel09}.

Regarding the stellar population component our results point 
to a mean metallicity solar to above solar, if we consider the  
light-weighted values,  while for the mass-weighted mean metallicity 
our results indicate a sub-solar value. We associate this discrepancy with the well 
known age-metallicity degeneracy, i.e. for a fixed mass, a high-metallicity stellar population looks cooler - and older - 
than a low-metallicity SP, thus resulting in a higher $M/L$ ratio.  Moreover, this is consistent 
with a galaxy chemical enrichment scenario in which the young population is enriched by the evolution of the 
early massive stars. In this context, the light-weighted metallicity is more sensitive to the 
young component, while the mass-weighted metallicity to the old stellar population.

\section{Conclusions}

\begin{figure*}
\begin{minipage}[b]{0.45\linewidth}
\includegraphics[width=\textwidth]{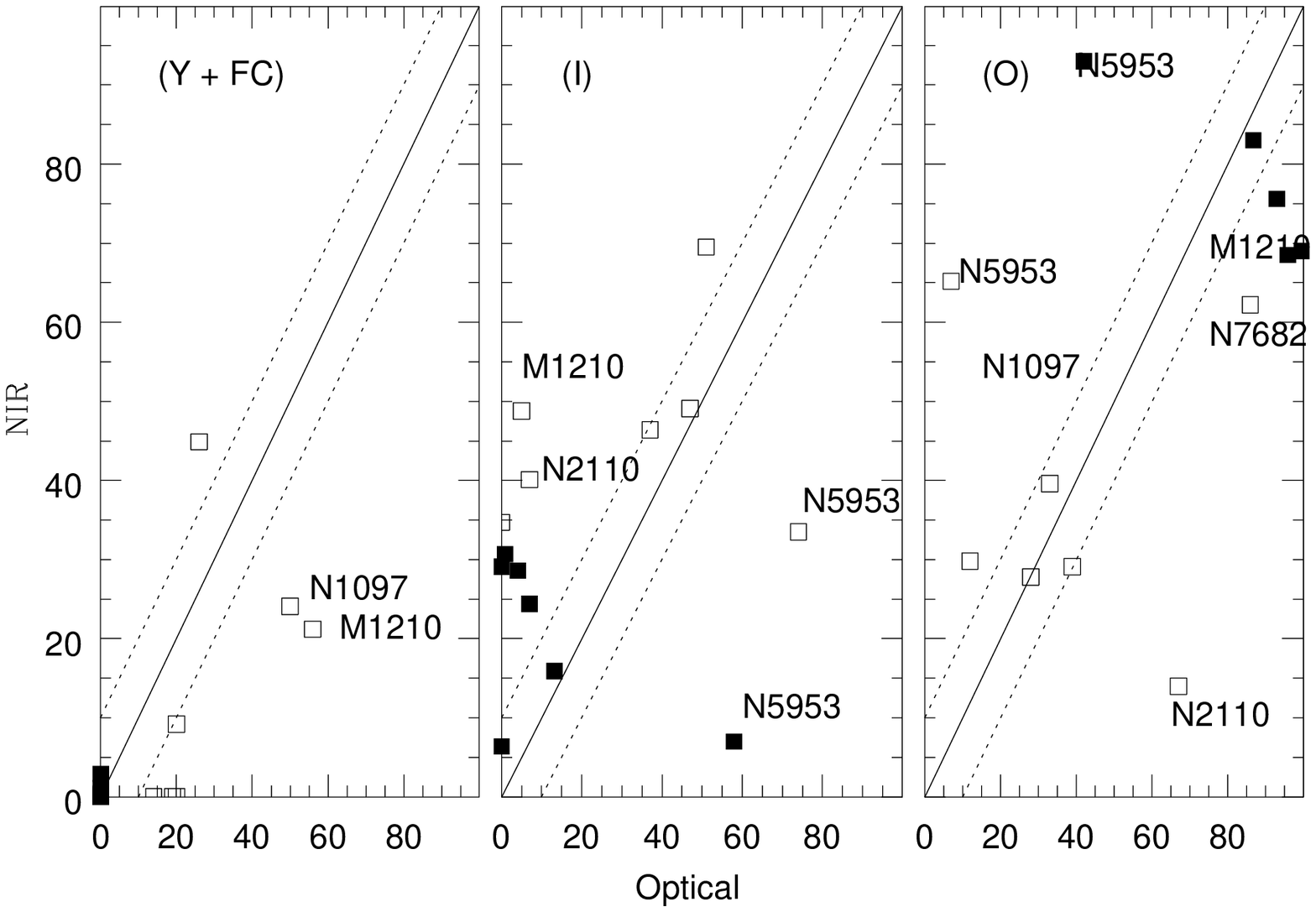}
\vspace{-2.5cm}
\caption{Comparison of the population vectors obtained in the NIR (this work) and in the optical (CF04), for 7 objects in common.
The symbols indicate the population vectors. The full line is the identity line, the dotted lines represent $\pm$10\% 
deviation from the identity. Open and filled symbols are the flux and mass fraction, respectively.  }
\label{compara}
\end{minipage}\hfill
\begin{minipage}[b]{0.45\linewidth}
\includegraphics[width=\textwidth]{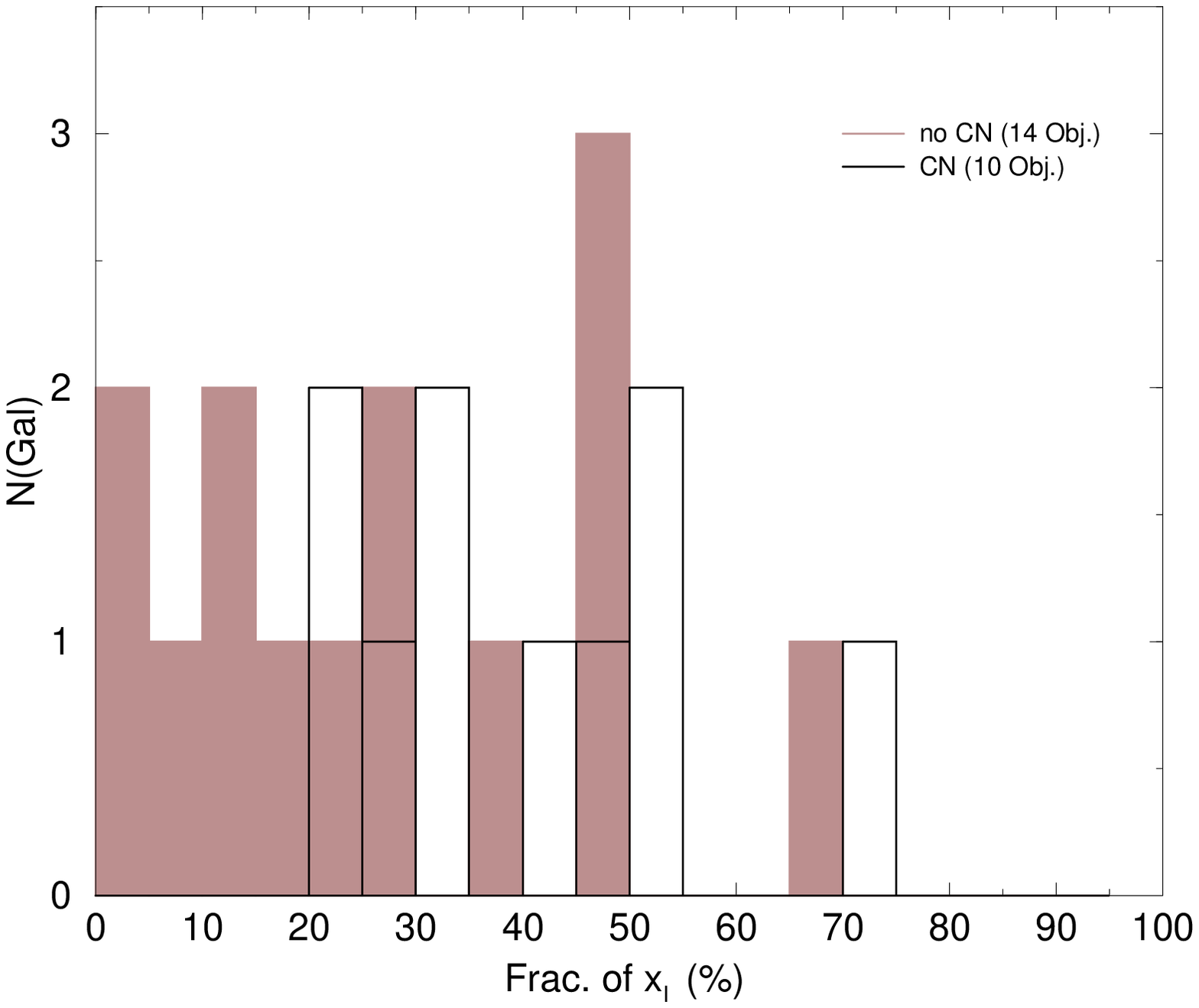}
\caption{Histograms comparing the intermediate age component of the galaxies with CN detection (empty histogram) 
and un-detection (shaded).}
\label{cn}
\end{minipage}\hfill
\end{figure*}

The main results can be summarised as follows:
\begin{itemize}
\item We found evidence of correlation among
the W$_{\lambda}$ of  Si\,I\,1.59$\mu$m $\times$ Mg\,I\,1.58$\mu$m, equally for both kinds of activity. Part of the
$W\rm_{Na\,I\,2.21\mu m}$ and $W\rm_{CO\,2.3\mu m}$ strengths and the correlation between 
$W\rm_{Na\,I\,2.21\mu m}$ and $W\rm_{Mg\,I\,1.58\mu m}$ appears to be accounted for 
by galaxy inclination. 

\item For the 7 objects in
common with previous optical studies (based on the same method of analyses), 
the NIR stellar population synthesis does not reproduce well
the optical results (see Fig.~\ref{compara}). 

\item Our synthesis shows significant differences between Sy\,1 and Sy\,2
galaxies. The hot dust component is required to fit the $K$-band spectra of $\sim$80\%
of the Sy\,1 galaxies, and only of $\sim$40\% of the Sy\,2. Besides, about 50\% of the Sy\,2 galaxies
require a featureless component contribution in excess of 20\%, while this fraction increases
to about 60\% in the Sy\,1. Also, in about 50\% of the Sy\,2, the combined FC and $X_Y$ components
contribute with more than 20\%, while this occurs in 90\% of the Sy\,1. This suggests recent
star formation (\cite[Cid Fernandes et al. 2005]{cid05}) in the central region of our galaxy sample. 

\item We found that the light at 1.223$\mu$m in central regions of the galaxies studied here
contain a substantial fraction of intermediate-age SPs with a mean metallicity near solar. Moreover, our analysis 
confirms that the 1.1$\mu$m CN band can be taken 
as an unambiguous tracer of intermediate-age stellar populations.

\end{itemize}

What emerges from this work is that the NIR may be taken as an excellent window to study the stellar
population of Sy\,1 galaxies, as opposed to the usually heavily attenuated optical range. Our
approach opens a new way to investigate and quantify the contribution of the three most important
NIR continuum components observed in AGNs. This new thecnique is described in 
\cite[Riffel et al. 2009]{riffel09}.

\end{document}